\documentclass[journal]{IEEEtran}

\usepackage{cite}
\usepackage{graphicx}
\usepackage{amsmath,amssymb}
\usepackage{array}
\usepackage{booktabs}
\usepackage{algorithm}
\usepackage{algpseudocode}
\usepackage{placeins}
\usepackage{capt-of}
\usepackage{url}

\title{DynaTE: Accelerating Diffusion LLMs via Dynamic Token Execution}
\author{
	Minghan Jiang,
	Jiayi Wang,
    Shuaiting Li,
	Haibin Shen,
	Kejie Huang\vspace{-3mm}

    \thanks{M. Jiang, J. Wang, S. Li, H. Shen, and K. Huang are with the College of Information Science \& Electronic Engineering, Zhejiang University, Hangzhou 310027, China, email: jiangmh@zju.edu.cn; wangjy\_zju@zju.edu.cn; list@zju.edu.cn;  shen\_hb@zju.edu.cn; huangkejie@zju.edu.cn. (corresponding author: K. Huang)}
    \thanks{This work has been submitted to IEEE for possible publication. Copyright may be transferred without notice, after which this version may no longer be accessible.}
}

\begin{document}
\maketitle

\begin{abstract}
Diffusion-based LLMs (dLLMs) have recently emerged as a promising alternative to autoregressive (AR) LLMs by enabling bidirectional parallel refinement, alleviating the sequential decoding bottleneck of AR generation. However, their parallel iterative refinement mismatches AR accelerators optimized for sequential decoding and their discrete token generation differs from DiT accelerators designed for continuous denoising. Recent dLLM accelerators have explored workload-specific optimizations to reduce vocabulary processing overhead and redundant computation across denoising iterations. However, these approaches retain all tokens in parallel execution, despite varying token refinement utility and execution requirements.

This paper presents DynaTE, a hardware--software co-design architecture that dynamically adapts accelerator execution to evolving token states during dLLM decoding. DynaTE first enables adaptive token execution by skipping low-utility token computation, while a dimension-reconfigurable PE array maintains high utilization under varying active-token patterns. Second, DynaTE exploits dynamic token dependencies through FLDD to refine a small number of locally dependent tokens within the current iteration, reducing the overall number of denoising iterations, while a Merge--Split--Merge dataflow hides the resulting serial overhead. Third, a streaming vocabulary engine interleaves multiple token streams from the LM head to accommodate irregular output variations caused by selective token computation and uneven vocabulary-selection demands.

Evaluated on two representative dLLMs, DynaTE achieves 2.05--2.78$\times$ speedup and 2.99--3.93$\times$ higher energy efficiency over state-of-the-art dLLM accelerators, while delivering 2.55$\times$ speedup and 6.07$\times$ higher energy efficiency over Jetson AGX Orin.
\end{abstract}


\section{Introduction}
Large language models (LLMs) have achieved remarkable success with the autoregressive (AR) generation paradigm~\cite{ge2023openagi,du2024survey,wu2024chateda,zhao2026codev}. However, AR decoding suffers from limited parallelism due to the sequential dependency among generated tokens, resulting in increasing inference latency for large-scale models~\cite{yan2025decoding,chen2024understanding}.

\begin{figure}[!t]
  \centering
  \includegraphics[width=\linewidth]{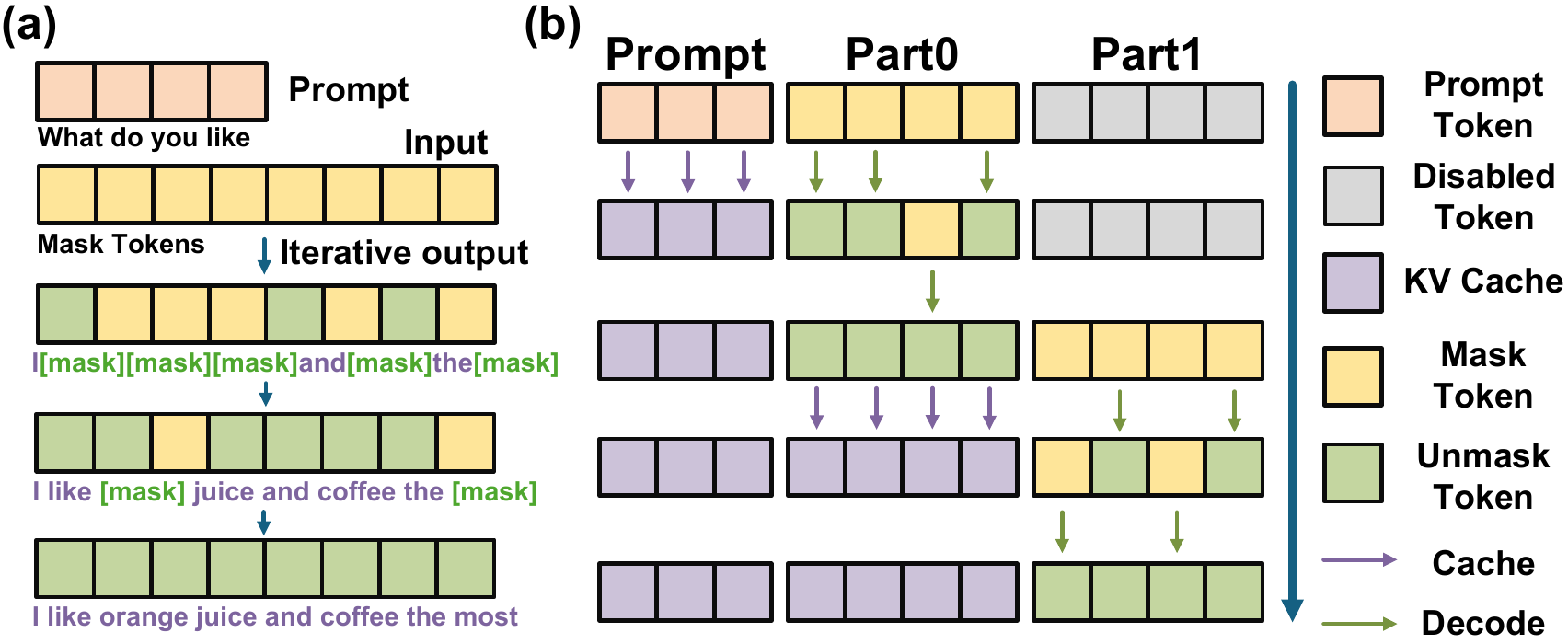}
  \caption{(a) Iterative decoding process of a dLLM. (b) Fast-dLLM's part-wise decoding with approximate KV-cache reuse.}
  \label{fig:dLLM_vs_AR}
\end{figure}

Recently, diffusion large language models (dLLMs) have emerged as an alternative generation paradigm by formulating text generation as an iterative denoising process~\cite{nie2025large,ye2025dream,gong2025scaling}. Unlike AR models that generate tokens sequentially, dLLMs simultaneously refine multiple masked tokens and progressively recover the complete sequence through multiple denoising iterations~\cite{sahoo2024simple,shi2024simplified}, as shown in Fig.~\ref{fig:dLLM_vs_AR}(a). Fast-dLLM further introduces part-wise decoding with approximate KV-cache reuse~\cite{wu2025fast}, as shown in Fig.~\ref{fig:dLLM_vs_AR}(b). This fundamentally different generation process introduces new execution characteristics for accelerator design.

However, existing accelerators are not directly optimized for such workloads. AR-oriented accelerators typically assume sequential single-token generation~\cite{chen2024understanding,huang2025edgellm,li2025hummingbird,guan2026aptqplus,zheng2026terafly}, while diffusion accelerators such as DiT accelerators mainly target dense updates over continuous latent representations~\cite{kim2025mixdit}. In contrast, dLLM inference operates on discrete token spaces and requires vocabulary projection and sampling during each denoising iteration, leading to distinct computation and data movement patterns~\cite{nie2025large,lou2026npudesigndiffusionlanguage}.

Recently, several accelerators have been proposed specifically for dLLM inference. DART focuses on accelerating the vocabulary sampling stage by introducing dedicated hardware support for large-vocabulary processing~\cite{lou2026npudesigndiffusionlanguage}, while dLLM-OPU exploits the similarity across denoising iterations through temporal activation reuse~\cite{wei2026dllm}. These works demonstrate the importance of adapting accelerator designs to dLLM-specific workloads. However, these designs retain all tokens in parallel execution during denoising, underexploring the varying importance and execution requirements among tokens. Exploiting such token-level variations requires dynamically adapting computation and handling the resulting irregular execution patterns.

\begin{figure}[t]
  \centering
  \includegraphics[width=1\linewidth]{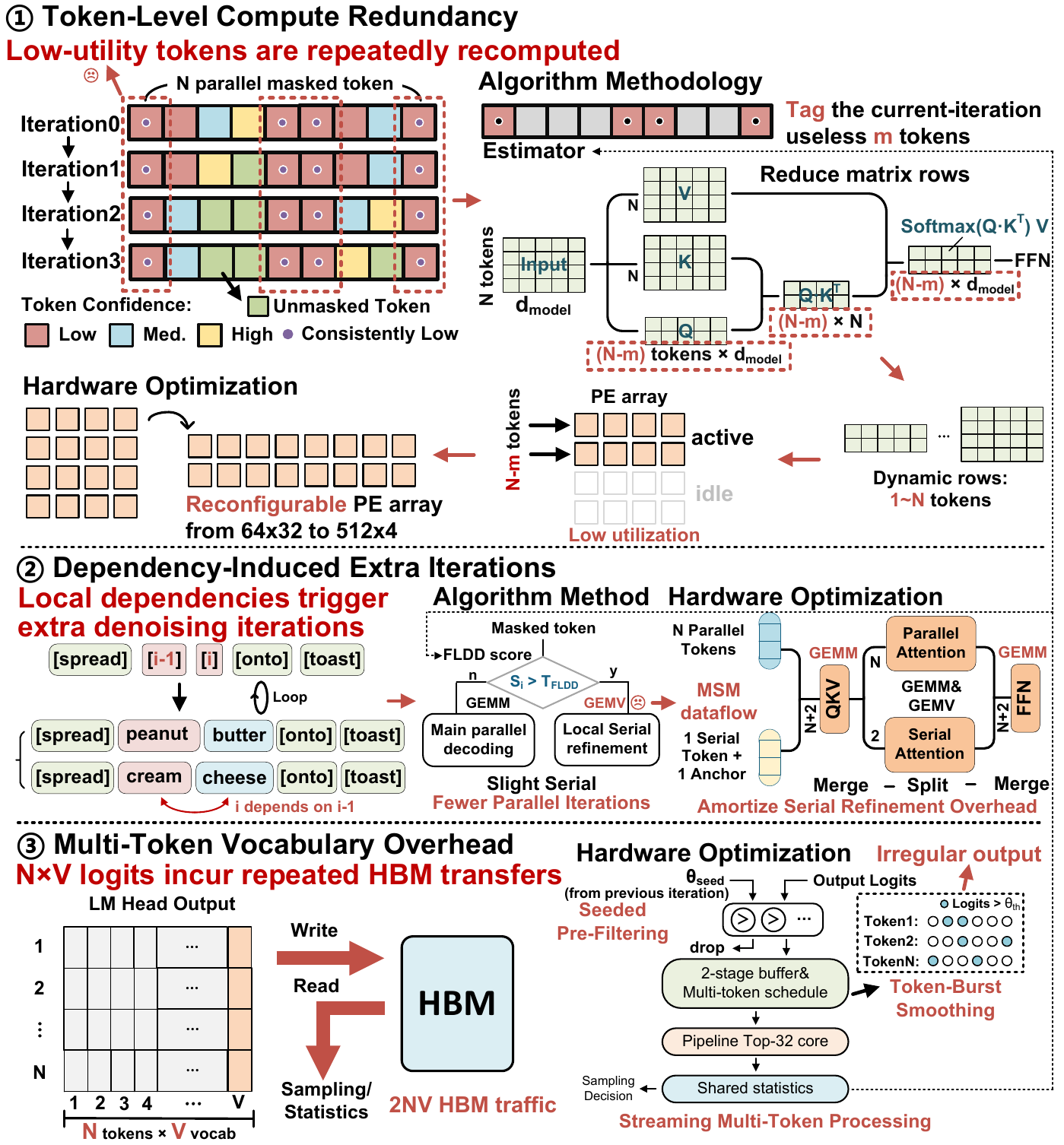}
  \caption{Three inefficiencies in dLLM decoding and DynaTE's corresponding algorithm--hardware optimizations.}
  \label{fig:dllm_difficulty_intro}
\end{figure}

To investigate this issue, we characterize dLLM workloads and observe that token execution requirements continuously change during denoising. Specifically, different tokens may require different computation resources, exhibit different refinement dependencies, and generate different vocabulary processing demands. Such token-level execution dynamics introduce new challenges for accelerator design. As shown in Fig.~\ref{fig:dllm_difficulty_intro}, we identify three manifestations of token-level execution dynamics, spanning iterative Transformer computation and vocabulary-side processing.

\textit{Challenge 1 (Dynamic Token Participation):}
During dLLM denoising, different tokens exhibit different refinement utilities across iterations. Although all unresolved tokens are repeatedly processed, many low-utility tokens provide limited improvement to generation quality while still incurring complete Transformer computation and data movement. Therefore, efficient dLLM accelerators should dynamically adjust token participation according to their refinement utility.

\textit{Challenge 2 (Dynamic Token Dependency):}
Although dLLMs decode multiple tokens in parallel, some locally coupled tokens cannot be resolved independently. Their updated context is only available in later iterations, causing repeated revisions and additional sequence-wide denoising steps.

\textit{Challenge 3 (Dynamic Vocabulary Execution):}
Each denoising iteration generates vocabulary distributions for multiple token positions. However, vocabulary outputs exhibit irregular generation patterns across token streams, leading to inefficient utilization of conventional fixed-granularity vocabulary selection engines. Moreover, materializing the complete N$\times$V logit tensor introduces substantial data movement between the LM head and sampling stage.

To address these challenges, we propose DynaTE, a dynamic token execution architecture for diffusion LLM inference. DynaTE adapts accelerator execution according to evolving token states through three key architectural mechanisms.

First, DynaTE addresses dynamic token participation through adaptive computation execution.
It introduces a \textbf{token utility-aware execution} mechanism to identify tokens with limited refinement benefits and avoid unnecessary Transformer computation. To efficiently support varying numbers of active tokens, DynaTE employs a \textbf{dimension-reconfigurable} PE array that dynamically adjusts execution granularity between token and model dimensions, improving hardware utilization under different participation patterns.

Second, DynaTE addresses dynamic token dependency through dependency-aware intra-iteration refinement.
Instead of resolving transient token dependencies through additional global denoising iterations, DynaTE employs \textbf{Fast Local Dependency Detection (FLDD)} to identify refinement opportunities and selectively performs local refinement within the current iteration. A \textbf{Merge-Split-Merge (MSM)} dataflow is designed to integrate refinement operations with regular parallel execution while minimizing serialization overhead.

Third, DynaTE addresses dynamic vocabulary execution through streaming multi-token processing.
DynaTE introduces a \textbf{streaming multi-token vocabulary engine} that directly processes vocabulary outputs as token streams. By exploiting irregular output patterns among different token positions, DynaTE improves vocabulary selection efficiency and reduces unnecessary data movement during multi-token generation.

DynaTE further incorporates iteration-adaptive activation precision with a shared splittable MAC datapath to reduce computation cost during decoding.

The main contributions of this work are summarized as follows:
\begin{itemize}
  \item We propose a \textbf{dynamic token execution approach} that adapts token participation and selective local refinement to evolving token states. Token-wise query/FFN pruning reduces low-utility computation, while Fast Local Dependency Detection (FLDD) guides local refinement to reduce unnecessary denoising iterations.

  \item We develop an architecture that \textbf{supports the resulting variable workloads} through dimension-reconfigurable computation and an MSM dataflow. A streaming multi-token vocabulary engine adapts to irregular token-wise LM-head outputs during denoising and supplies shared distribution statistics for sampling, refinement, and subsequent pruning while avoiding full-logit HBM transfers.

  \item We evaluate DynaTE on LLaDA-8B and Dream-7B across five benchmarks. The evaluated system achieves \textbf{2.05--2.78$\times$} speedup and \textbf{2.99--3.93$\times$} higher energy efficiency over the evaluated prior dLLM accelerators, and \textbf{2.55$\times$} speedup and \textbf{6.07$\times$} higher energy efficiency over Jetson AGX Orin.
\end{itemize}

\section{Background \& Motivation}
\subsection{dLLM basics}
Diffusion large language models (dLLMs) reformulate text generation as an iterative denoising process rather than the conventional left-to-right autoregressive (AR) prediction \cite{nie2025large}. Given an original token sequence $x_0 = (x_1, \dots, x_L)$, the forward \emph{masking} process gradually replaces each token with a special \texttt{[MASK]} symbol according to a masking ratio $t \in [0,1]$:
\begin{equation}
q_t(x_t|x_0) = \prod_{i=1}^{L} \text{Cat}\!\big(x_i^t; (1-t)\,\delta_{x_i^0} + t\,\delta_{\texttt{[MASK]}}\big),
\end{equation}
where $x_t$ denotes the partially masked sequence at time $t$.
As shown in Fig.~\ref{fig:dLLM_vs_AR}(a), during inference, the model reconstructs the text through multiple denoising iterations, where all tokens are re-computed in parallel as
\begin{equation}
\hat{x}^{(k)} = \arg\max_{x} p_\theta(x|x_t^{(k)}),
\end{equation}
and tokens with low confidence are re-masked for regeneration, i.e.,
\begin{equation}
x_t^{(k+1)} = \text{Remask}\!\left(\hat{x}^{(k)}, \tau\right).
\end{equation}
This iterative re-masking mechanism, as illustrated in Fig.~\ref{fig:dLLM_vs_AR}(a), enables bidirectional context modeling and parallel decoding, effectively alleviating the latency bottleneck of AR models.

Building upon the LLaDA-dLLM framework, Fast-dLLM introduces a part-wise approximate KV caching mechanism, as depicted in Fig.~\ref{fig:dLLM_vs_AR}(b), which reuses attention activations across adjacent diffusion steps \cite{wu2025fast}. By exploiting the high temporal similarity of KV states between iterations, this method achieves substantial computational reuse with negligible degradation in generation quality.

Beyond the savings from part-wise decoding and KV-cache reuse, token-level variations in refinement utility and local dependencies offer further opportunities to reduce computation during denoising.

\subsection{Execution Mismatch with Existing Accelerators}
\label{sec:mismatch}
The mismatch stems from the execution unit around which each accelerator is designed.

AR accelerators treat one newly generated token as the basic unit of decoding~\cite{zheng2026lora,ma2026aptllm}. Their datapaths are optimized for low-parallelism execution, while their cache systems assume that new states are appended without revisiting earlier positions. A dLLM instead updates several positions within the same block and may revisit them over multiple iterations. Its active workload therefore changes during generation, making a fixed single-token execution path inefficient.

Accelerators for continuous-latent DiTs are built around a different form of iteration~\cite{li2026aadit}. They retain a fixed spatial-token layout and densely update the latent representation at every timestep. The output remains a compact continuous tensor that is consumed by a numerical scheduler. In contrast, a dLLM produces a vocabulary distribution for each active position and then changes its discrete token state through commit or remask decisions. Although the Transformer compute units remain reusable, the downstream dataflow does not match vocabulary-wide processing or token-state control.

\subsection{Workload Characterization and Design Opportunities}
\label{sec:motivation}

\begin{figure}[!t]
  \centering
  \includegraphics[width=\linewidth]{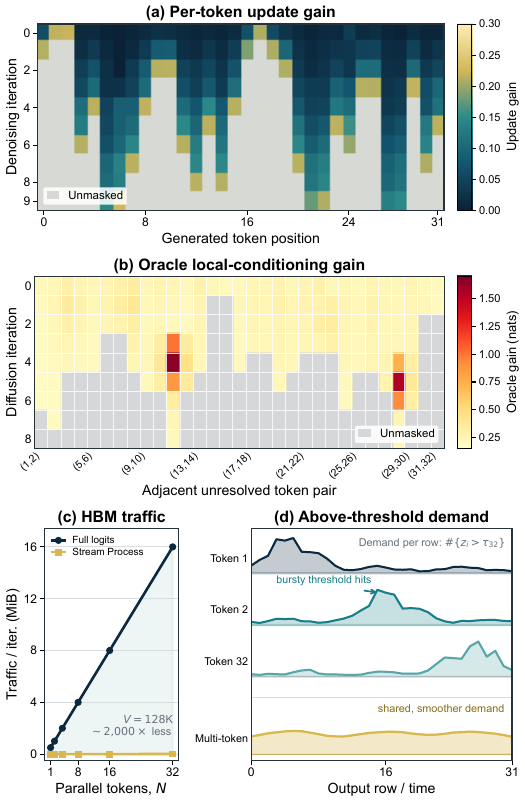}
  \caption{Workload characterization of the three optimization
  opportunities: (a) per-token update gain across denoising
  iterations; (b) oracle local-conditioning gain between adjacent
  unresolved tokens; (c) full-logit HBM traffic; and
  (d) bursty per-token versus smoothed multi-token processing demand.}
  \label{fig:motivation}
\end{figure}

Section~\ref{sec:mismatch} explains why existing accelerators are
poorly matched to dLLM execution. We further characterize the
inefficiencies that remain in iterative decoding.
Fig.~\ref{fig:motivation} presents two representative decoding traces
and the vocabulary-side data movement. These observations correspond
to the three challenges introduced in
Fig.~\ref{fig:dllm_difficulty_intro}.

\noindent\textbf{Observation 1: Token-update utility is uneven.}
A dLLM repeatedly processes unresolved positions over multiple
denoising iterations. However, recomputing a token does not always
produce a meaningful improvement. To reveal this behavior, we define
the oracle update gain of token $i$ at iteration $t$ as
\begin{equation}
G_i^{(t)}
=
\left[
\log p_{i,\mathrm{after}}^{(t)}(y_i)
-
\log p_{i,\mathrm{before}}^{(t)}(y_i)
\right]_{+},
\label{eq:update_gain}
\end{equation}
where $y_i$ denotes the reference token and
$[x]_{+}=\max(x,0)$.

As shown in Fig.~\ref{fig:motivation}(a), many unresolved positions
receive little benefit from an update, even though they still pass
through the complete Transformer. The high-gain positions also change
across iterations. Therefore, this redundancy cannot be captured by a
fixed token mask. An efficient accelerator should identify low-utility
positions at runtime and adapt the active computation accordingly.

\noindent\textbf{Observation 2: Strong local dependencies are sparse
and transient.}
Parallel decoding assumes that unresolved positions can be refined
independently within the same iteration. This assumption does not
always hold. Resolving one token may provide critical context for a
nearby token, making a short serial refinement more effective than
another fully parallel iteration.

For an adjacent unresolved pair $(i,j)$, we define the directional
coupling gain as
\begin{equation}
C_{i\rightarrow j}^{(t)}
=
\left[
\log p_j^{(t)}(y_j \mid x_i=y_i)
-
\log p_j^{(t)}(y_j \mid x_i\ \mathrm{unresolved})
\right]_{+}.
\label{eq:coupling_gain}
\end{equation}

The reference tokens are again used only to expose the underlying
opportunity.

Fig.~\ref{fig:motivation}(b) shows that high coupling gains appear
only at isolated token pairs and iterations. Their locations also
change during decoding. Serializing the entire block would therefore
sacrifice parallelism for positions that do not benefit from ordered
updates. Instead, serial refinement should be restricted to the small
number of temporary dependency hotspots.

\noindent\textbf{Observation 3: Vocabulary outputs incur substantial
off-chip traffic.}

Each denoising iteration projects $N$ active token states onto a
vocabulary of size $V$, producing an $N\times V$ logit tensor. In a
decoupled LM-head and sampling pipeline, this tensor is first written to
HBM and later read back for vocabulary processing. If each logit occupies
$b$ bytes, the intermediate traffic per iteration is

\begin{equation}
D_{\mathrm{logit}} = 2NVb,
\end{equation}

where the factor of two accounts for one write and one read. As shown in
Fig.~\ref{fig:motivation}(c), this traffic increases linearly with the number of tokens
decoded in parallel and is repeatedly incurred across denoising
iterations.

Meanwhile, the numbers of logits exceeding the current Top-32 threshold
are highly uneven over time. As shown in Fig.~\ref{fig:motivation}(d), a single token stream
exhibits bursty candidate arrivals, while the bursts of different tokens
often occur at different vocabulary positions. Processing token streams
independently therefore leaves the vocabulary-selection pipeline
underutilized, whereas interleaving multiple token streams can smooth the
aggregate update demand.

The complete logit tensor need not remain available after generation.
Leading vocabulary candidates and full-vocabulary normalization
statistics can instead be extracted directly from the LM-head output
stream. Therefore, vocabulary outputs should be summarized before
crossing the HBM boundary, while multi-token interleaving should be used
to smooth irregular candidate arrivals and improve shared selection
hardware utilization.

\section{Algorithm Methodology}

\subsection{Token-wise Query-Pruning Strategy}
\label{prun}

DynaTE first uses the distribution statistics from the previous
iteration and the positions of decoded tokens to identify masked
tokens with low expected refinement utility. Within the current
decoding block $\mathcal{B}^{(r)}$, the pruning candidate set
is defined as
\begin{equation}
\begin{aligned}
\mathcal{C}^{(r)}
=
\bigl\{ i \in \mathcal{B}^{(r)} \,\bigm|\,
&x_i^{(r)}=[\mathrm{MASK}],
\ H_i^{32,(r-1)}\geq\tau_H, \\
&C_i^{(r-1)}\leq\tau_C,
\ D_i^{(r)}\geq\tau_D
\bigr\}.
\end{aligned}
\end{equation}
Here, $H_i^{32,(r-1)}$ denotes the Top-$32$ truncated entropy,
$C_i^{(r-1)}$ is the token confidence, and $D_i^{(r)}$ is the
distance to the nearest decoded token.

\begin{figure}[!t]
  \centering
  \includegraphics[width=\linewidth]{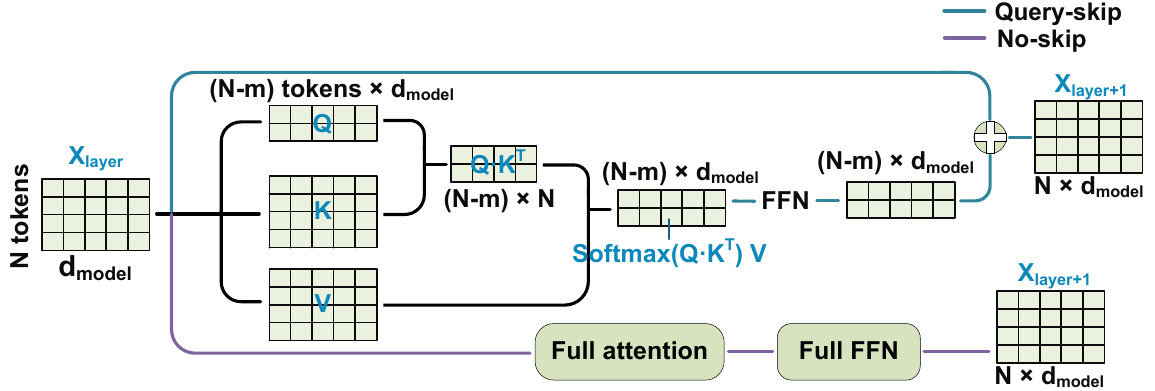}
  \caption{Token-wise pruning dataflow. Query-skip removes selected
  tokens from query-side attention and FFN computation while retaining
  their keys and values for the remaining tokens; no-skip executes the
  complete layer.}
  \label{fig:prun}
\end{figure}

Candidate tokens may exhibit different contextual contributions across
Transformer layers. DynaTE therefore accumulates the average attention
mass received by candidate token $i$ at layer $\ell$:

\begin{equation}
M_{\ell,i}
=
\frac{1}{N_h|\mathcal{Q}_{\ell}|}
\sum_{h=1}^{N_h}
\sum_{q\in\mathcal{Q}_{\ell}}
A_{\ell,h,q,i},
\end{equation}

and cumulatively updates the next layer-wise pruning set as

\begin{equation}
P_{\ell+1}^{(r)}
=
P_{\ell}^{(r)} \cup
\left\{
i \in \mathcal{C}^{(r)} \setminus P_{\ell}^{(r)}
\;\middle|\;
M_{\ell,i} < T_{\ell+1}
\right\}.
\end{equation}

where $T_{\ell+1}$ is an offline-calibrated threshold and
$P_1^{(r)}=\varnothing$ for each decoding iteration $r$.
Thus, a token is newly selected for pruning in the next layer only when
it is predicted to have low refinement utility and receives little
attention from the active queries, with the received attention mass
serving as a layer-wise safeguard against pruning contextually
influential tokens. The attention mass
is accumulated while the normalized attention probabilities are
streamed into the $AV$ computation, requiring only lightweight
accumulation without storing the complete attention matrix or
introducing additional off-chip accesses. As shown in
Fig.~\ref{fig:prun}, for a pruned token, the hidden state is passed
unchanged to the next layer through an identity bypass. Its keys and
values are recomputed at each layer from the normalized layer input
and remain accessible to active tokens, while its query-side attention
and FFN computations are skipped. Pruning persists through the
remaining layers of the current decoding iteration and is reassessed
in the next iteration.

\subsection{Fast Local Dependency Detection for Selective Serial Refinement}
\label{fldd}

Since all tokens in parallel decoding are predicted from the same
pre-update context, a newly resolved token cannot immediately affect
its neighboring positions within the current iteration. DynaTE
introduces Fast Local Dependency Detection (FLDD), which compares
cross-iteration distribution changes at neighboring positions to
identify anchors for local serial refinement.

FLDD measures the difference between the cross-iteration Top-$32$
truncated KL divergences of neighboring positions $i$ and $i+1$:

\begin{equation}
\begin{aligned}
E_i^{(r)}=\Bigl|&
D_{\mathrm{KL32}}\!\left(p_i^{(r)}\Vert p_i^{(r-1)}\right)\\
&-D_{\mathrm{KL32}}\!\left(p_{i+1}^{(r)}\Vert p_{i+1}^{(r-1)}\right)
\Bigr|.
\end{aligned}
\end{equation}

A smaller $E_i^{(r)}$ indicates more similar magnitudes of
cross-iteration distribution change, suggesting a potential local
dependency.

FLDD then compares the Top-$K$ candidate sets of position $i$ across
two consecutive iterations:

\begin{equation}
O_i^{(r)}
=
\frac{
\left|
\operatorname{TopK}\!\left(p_i^{(r-1)}\right)
\cap
\operatorname{TopK}\!\left(p_i^{(r)}\right)
\right|
}{K},
\end{equation}

where $K=4$ in our design. A small overlap indicates that the dominant
candidates of this position have changed substantially in the current
iteration.

Finally, FLDD uses the latest available Top-$32$ truncated entropy
$H_i^{32,(r)}$ to characterize the concentration of the prediction
distribution. For a position whose confidence remains below the
commitment threshold, lower entropy indicates that probability mass
is concentrated among fewer candidates, even though the position
is not yet confidently resolved. This metric describes current
candidate concentration rather than cross-iteration stability.
The three metrics are combined into the local dependency score:

\begin{equation}
\begin{aligned}
S_i^{(r)}
={}&
\alpha_1
\left[1-\operatorname{Norm}\!\left(E_i^{(r)}\right)\right]
+\alpha_2\left(1-O_i^{(r)}\right) \\
&+\alpha_3\left[1-
\operatorname{Norm}\!\left(H_i^{32,(r)}\right)\right].
\end{aligned}
\end{equation}

When $S_i^{(r)}$ exceeds the FLDD threshold, position $i$ is selected
as an anchor. The neighboring token at position $i+1$ is refined
using the updated anchor's KV context, while the parallel path retains
the pre-anchor context.

FLDD aims to identify local refinement opportunities with low overhead,
reducing denoising iterations without requiring exhaustive dependency detection.

\subsection{Adaptive Activation Precision}
\label{adaptive}

The activation ranges of dLLMs vary with decoding progress and exhibit
different sensitivities across model modules. DynaTE therefore builds
calibration sets for different decoded-token ratios and offline profiles
the quantization sensitivity of individual modules, including QKV
projections, the LM head, and FFN projections. Based on this profiling,
a compact lookup table assigns W8A8 or W8A4 execution to each module at
different decoding stages. At runtime, the controller selects the
precomputed precision according to the current decoded-token ratio,
while a shared splittable MAC datapath executes either one W8A8 operation
or two W8A4 operations in parallel.

\section{Hardware Optimization}

\begin{figure}[!t]
  \centering
  \includegraphics[width=0.95\linewidth]{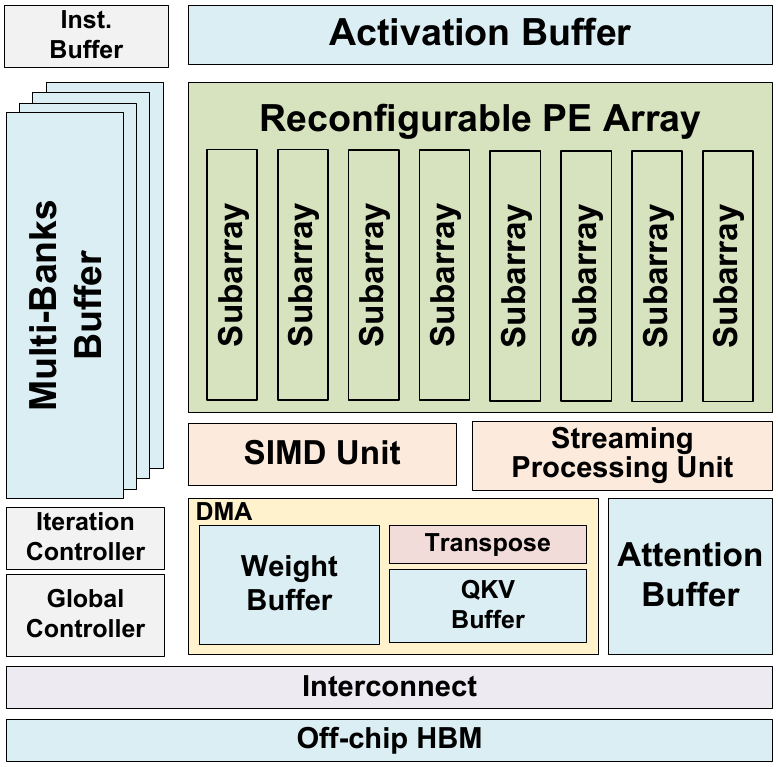}
  \caption{Overall architecture of DynaTE.}
  \label{fig:overall}
\end{figure}

Fig.~\ref{fig:overall} illustrates the overall architecture
of DynaTE, which comprises four main components: the computation
core, the memory hierarchy, the streaming processing unit, and
the control logic. The computation core includes a
dimension-reconfigurable PE array composed of eight subarrays,
a SIMD unit for vector operations, and a multi-bank buffer for
feeding the PE array. The memory hierarchy consists of the
Activation Buffer, Weight Buffer, QKV Buffer with an integrated
Transpose unit, and Attention Buffer for storing intermediate
results during MSM execution. The Streaming Processing Unit
directly processes LM-head outputs and generates Top-32 results
and distribution statistics. The Iteration Controller dynamically
configures token pruning, serial refinement, array shape, and
execution precision according to the current token states, while
the Global Controller manages the overall denoising flow.
Data are exchanged with off-chip HBM through the interconnect
for weight and KV-cache storage.

\subsection{Reconfigurable Computation Dimension}
\label{reconfigure}

\begin{figure*}[!t]
  \centering
  \includegraphics[width=\textwidth]{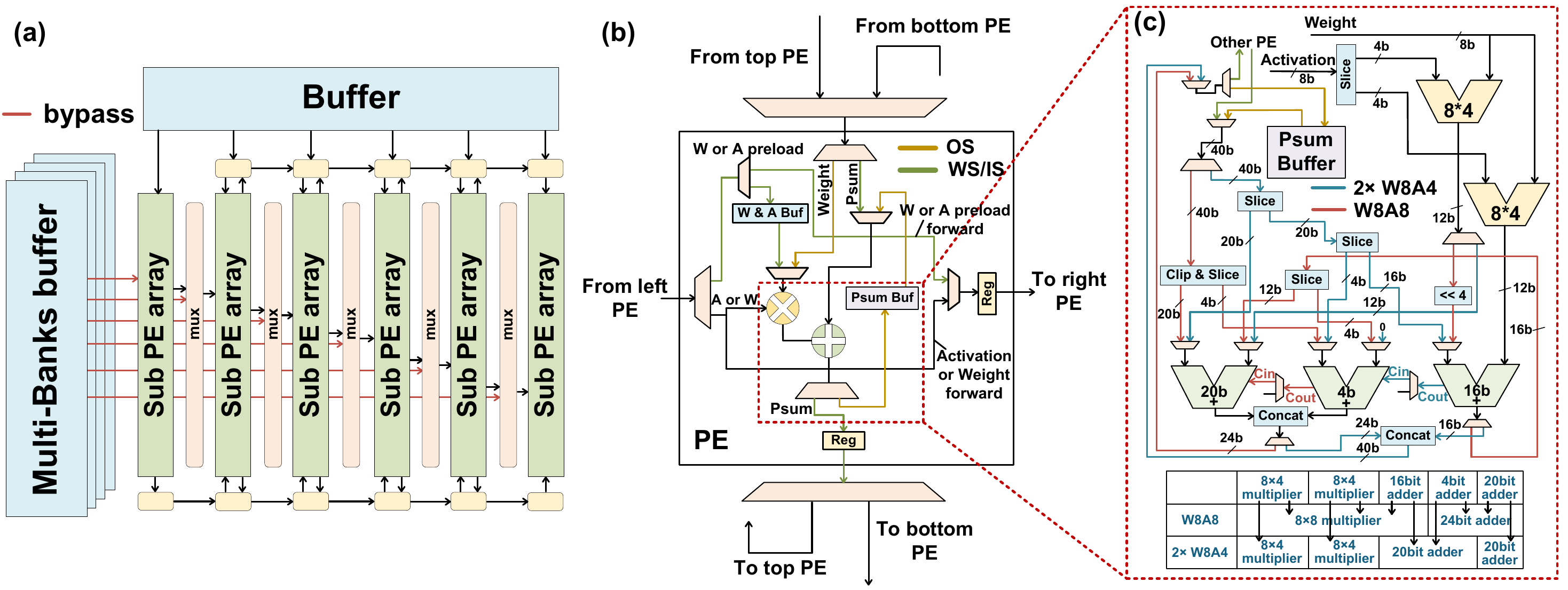}
  \caption{(a) Reconfigurable PE array, (b) internal PE structure, and
  (c) multi-precision splittable MAC unit.}
  \label{fig:array}
\end{figure*}

\begin{figure}[!t]
  \centering
  \includegraphics[width=\linewidth]{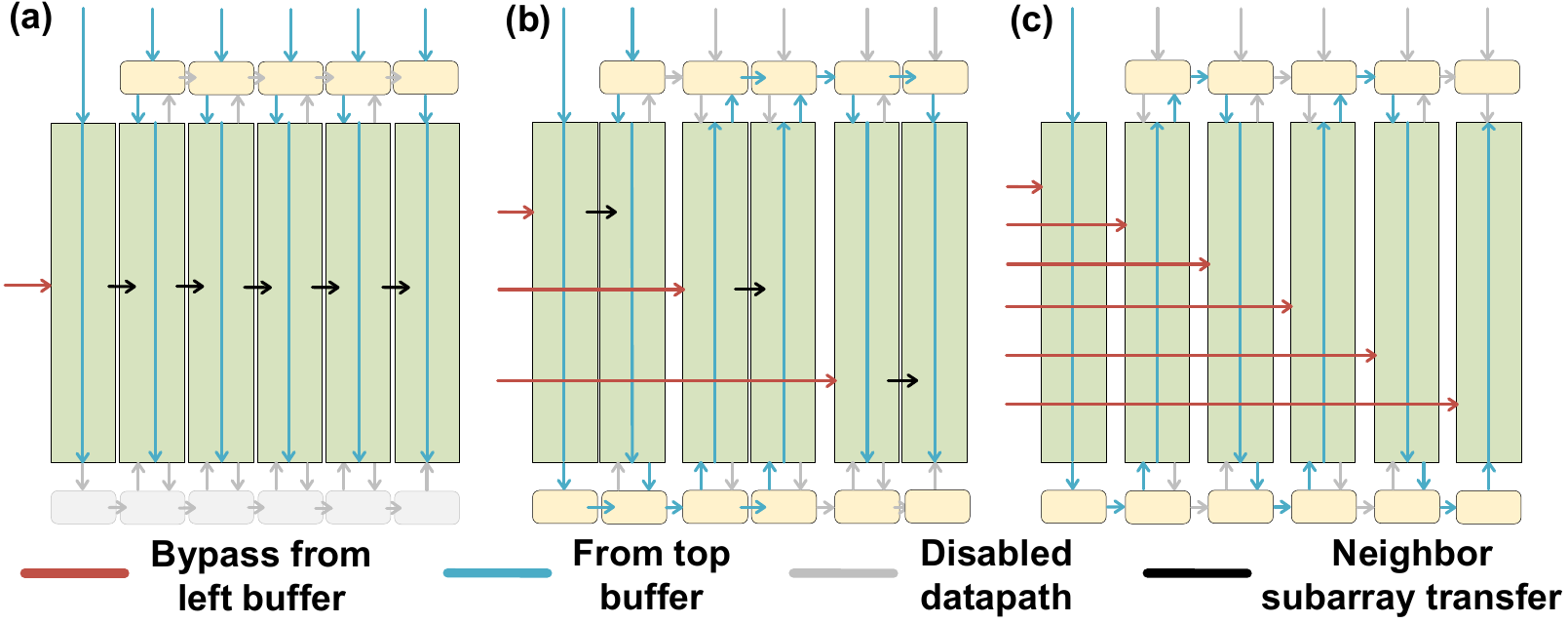}
  \caption{Representative PE-array reconfiguration modes:
  (a) canonical dataflow and (b)--(c) reshaped dataflows using
  left-buffer bypass and inter-subarray transfers.}
  \label{fig:reconfigure}
\end{figure}

Token pruning and serial refinement dynamically change the
effective matrix dimensions of Attention and FFN operations.
To maintain high utilization under these varying workloads,
we design a dimension-reconfigurable PE array composed of
eight $64\times4$ subarrays. With a fixed total of 2,048 MAC
units, the array supports $64\times32$, $128\times16$,
$256\times8$, and $512\times4$ configurations.

As shown in Fig.~\ref{fig:array}(a), adjacent
subarrays are connected through multiplexer-based gating circuits
that enable either horizontal propagation or direct bypass from
the left multi-bank buffer. Vertical data movement is controlled
by the routing logic above and below each subarray, with an
additional input path from the top buffer. As shown in
Fig.~\ref{fig:array}(b), each PE contains local
operand and partial-sum registers together with top and bottom
multiplexers, enabling bidirectional vertical routing and
supporting WS, IS, and OS dataflows.

Fig.~\ref{fig:reconfigure} illustrates representative array
configurations. In the canonical configuration, horizontal data
propagate through adjacent subarrays, while vertical inputs are
supplied from the top buffer. In reshaped configurations, the
left-buffer bypass and inter-subarray vertical paths are enabled
to combine multiple subarrays along the model dimension.
Adjacent subarray groups employ opposite propagation directions,
forming an S-shaped dataflow that reduces routing distance and
contention.

\subsection{Splittable MAC Unit}

Beyond array-level flexibility, each PE supports precision
reconfiguration to accommodate the stage-varying activation
precision of dLLMs. Conventional reconfigurable MAC designs
require two 20-bit accumulators for dual W8A4 execution, while
W8A8 execution requires a 24-bit accumulator, resulting in
redundant adder resources.

To reduce this overhead, we design a splittable MAC unit that
shares its multiplier and accumulator datapaths across precision
modes. As shown in Fig.~\ref{fig:array}(c), each
unit consists of two $8\times4$ multipliers, a 16-bit adder, a
4-bit adder, and a 20-bit adder.

\textit{W8A8 Mode:}
The 8-bit activation is divided into upper and lower 4-bit
slices and processed by the two $8\times4$ multipliers.
The two partial products are combined with the required shift
and sign extension to form an $8\times8$ product. The 20-bit
and 4-bit adders are concatenated to form a 24-bit accumulator.

\textit{Dual-W8A4 Mode:}
Two 4-bit activations are processed in parallel by the two
$8\times4$ multipliers. The 16-bit and 4-bit adders are
combined into one 20-bit accumulator, while the original
20-bit adder forms the other, enabling two independent W8A4
MAC operations.

This adder-sharing design allows each physical MAC to execute
either one W8A8 operation or two W8A4 operations without
duplicating the accumulator datapath.

\subsection{Streaming Multi-Token Vocabulary Processing Engine}

\begin{figure}[!t]
  \centering
  \includegraphics[width=\linewidth]{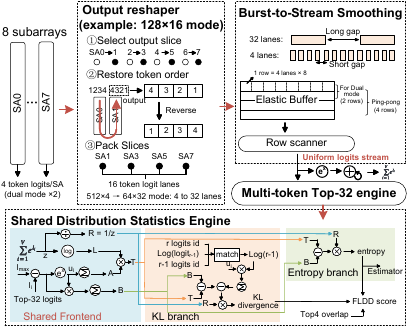}
  \caption{Streaming vocabulary-processing engine, including output
  reshaping, burst-to-stream smoothing, multi-token Top-32 selection,
  and shared distribution-statistics generation.}
  \label{fig:stream}
\end{figure}

To avoid transferring the complete $N\times V$ logit tensor between
the LM head and the sampling module, DynaTE directly connects a
streaming vocabulary-processing engine to the LM-head output. As shown
in Fig.~\ref{fig:stream}, the Output Reshaper selects valid outputs according to the
current PE-array configuration and restores the token order altered by
the S-shaped dataflow. The resulting $4$--$32$ logit lanes are packed
into a 32-bank elastic buffer and converted by the Row Scanner into a
uniform vocabulary stream. This design decouples the bursty outputs of
the dimension-reconfigurable PE array from the downstream processing
pipeline.

\begin{figure}[!t]
  \centering
  \includegraphics[width=\linewidth]{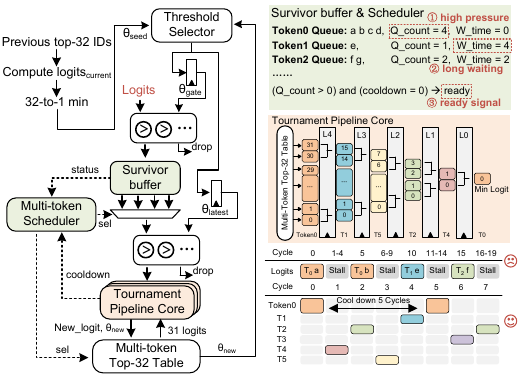}
  \caption{Multi-token Top-32 engine and its interleaved
  tournament-pipeline schedule.}
  \label{fig:top32}
\end{figure}

As shown in Fig.~\ref{fig:top32}, for each token, the Top-32 engine first retrieves
the Top-32 candidate IDs retained from the previous iteration and
computes their current logits. The minimum of these updated logits
initializes the selection threshold:

\begin{equation}
\theta_t^{\mathrm{seed}}
=
\min_{v\in\mathcal{T}_t^{(r-1)}}
z_t^{(r)}(v).
\end{equation}

During the subsequent full-vocabulary scan, logits below the threshold
are discarded, while the remaining candidates enter the comparison
pipeline. The Top-32 table is maintained as an unordered set, and each
update replaces only its current minimum. Since buffered candidates may
have passed an earlier threshold, they are revalidated against the
latest threshold before replacement. The warm start therefore reduces
comparison pressure while preserving exact Top-32 selection.

The numbers of surviving candidates from different token streams vary
substantially over time. DynaTE assigns a Survivor Buffer to each active
token and interleaves their candidates through shared Tournament
Pipeline Cores. The scheduler first serves tokens under high buffer pressure, then long-waiting tokens, and finally the remaining ready tokens in
a round-robin manner. Interleaving different token streams fills the
dependency interval between consecutive updates to the same Top-32
table, thereby smoothing bursty arrivals and improving pipeline
utilization. Algorithm~\ref{alg:multi_token_top32} summarizes this process.

\FloatBarrier
\begin{algorithm}[!t]
\caption{Multi-Token Top-32 Selection}
\label{alg:multi_token_top32}
\begin{algorithmic}[1]
\Require Previous Top-32 IDs $\mathcal{P}[t]$;
         streamed logits $(t,v,x)$
\Ensure Updated Top-32 table for each token $t$

\ForAll{tokens $t$}
    \State $\mathrm{Top32}[t]
        \gets \mathrm{CurrentLogits}\!\left[\mathcal{P}[t]\right]$
    \Comment{Reuse prior IDs to initialize the threshold}
    \State $\theta_{\mathrm{gate}}[t]
        \gets \min(\mathrm{Top32}[t])$
    \State $\theta_{\mathrm{latest}}[t]
        \gets \theta_{\mathrm{gate}}[t]$
\EndFor

\ForAll{incoming logits $(t,v,x)$}
    \If{$v \notin \mathcal{P}[t]$
        \textbf{and} $x > \theta_{\mathrm{gate}}[t]$}
        \State $\Call{Enqueue}{Q[t],(v,x)}$
    \EndIf
    \State Update cooldown and waiting-time counters

    \State $\mathit{ready}
        \gets (Q_{\mathrm{count}}>0)
        \land (\mathit{cooldown}=0)$
    \State $\mathit{high\_pressure}
        \gets \mathit{ready}
        \land (Q_{\mathrm{count}}\geq Q_{\mathrm{high}})$
    \State $\mathit{long\_waiting}
        \gets \mathit{ready}
        \land (\mathit{wait\_time}\geq W_{\mathrm{th}})$

    \If{$\Call{Any}{\mathit{high\_pressure}}$}
        \State $\mathit{mask}\gets\mathit{high\_pressure}$
    \ElsIf{$\Call{Any}{\mathit{long\_waiting}}$}
        \State $\mathit{mask}\gets\mathit{long\_waiting}$
    \Else
        \State $\mathit{mask}\gets\mathit{ready}$
    \EndIf

    \State $t^*\gets\Call{RRSelect}{\mathit{mask}}$
    \Comment{Interleave tokens to hide tree dependencies}

    \If{$t^*\neq\mathrm{INVALID}$}
        \State $(v^*,x^*)\gets\Call{Dequeue}{Q[t^*]}$

        \If{$x^*>\theta_{\mathrm{latest}}[t^*]$}
            \Comment{Revalidate stale-threshold candidates}
            \State $\Call{ReplaceTreeMin}{t^*,(v^*,x^*)}$
            \Comment{Replace only the current minimum}
            \State $\theta_{\mathrm{latest}}[t^*]
                \gets\Call{TreeMin}{t^*}$
            \State $\theta_{\mathrm{gate}}[t^*]
                \gets\theta_{\mathrm{latest}}[t^*]$
            \State $\mathit{cooldown}[t^*]\gets 5$
        \EndIf
    \EndIf
\EndFor

\While{$\Call{AnyQueueNonempty}{}$}
    \State Drain remaining survivors with the same policy
\EndWhile

\end{algorithmic}
\end{algorithm}

Meanwhile, the Shared Distribution Statistics Engine accumulates the
full-vocabulary Softmax normalization statistics and derives the
Top-32 truncated entropy, token confidence, cross-iteration KL
divergence, and Top-4 overlap from the retained candidates. The KL
branch compares neighboring positions' cross-iteration divergences
and supplies $1-\operatorname{Norm}(E_i^{(r)})$ to FLDD. The Top-32 results support
sampling, while the distribution statistics are retained for FLDD and
token pruning in the subsequent forward pass. The
streaming engine thus eliminates full-logit HBM transfers and unifies
sampling, dependency detection, and cross-iteration token evaluation
within a shared vocabulary-processing datapath.

\subsection{Merge-Split-Merge Dataflow}
\label{msm}

Serial refinement reduces the number of iterations required for token convergence, but introduces two critical inefficiencies if processed independently: (1) The refinement workload consists predominantly of general matrix-vector (GEMV) operations, which suffer from poor computational utilization on standard PE arrays optimized for matrix-matrix multiplication. (2) Serial processing of refinement causes the main parallel token generation to stall, leading to reduced overall throughput.

\begin{figure}[!t]
  \centering
  \includegraphics[width=0.95\linewidth]{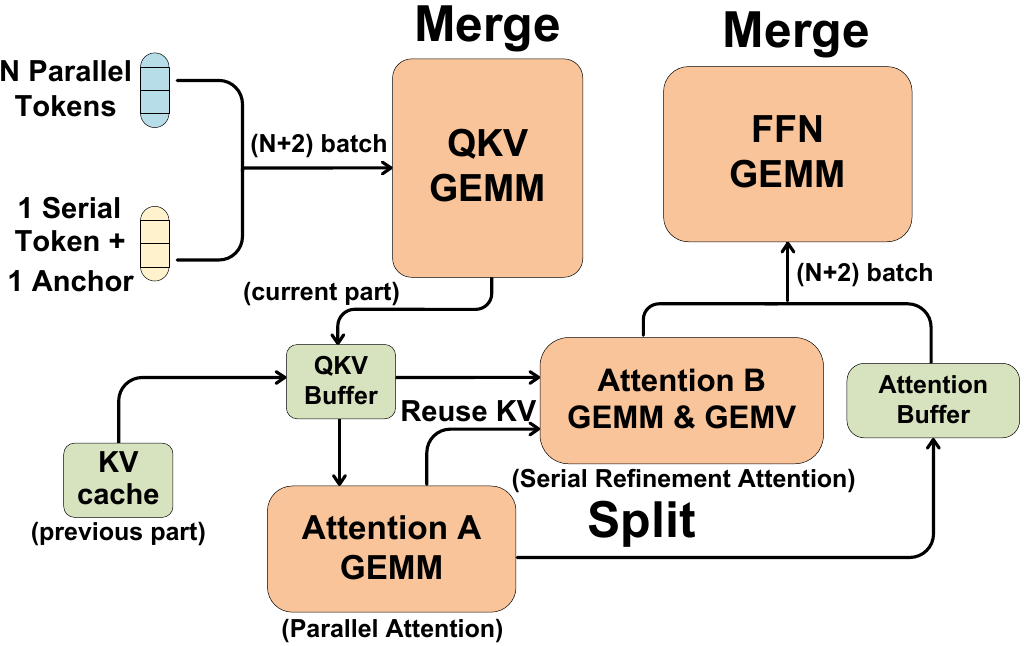}
  \caption{Merge--Split--Merge dataflow for integrating local serial
  refinement into parallel dLLM decoding.}
  \label{fig:merge}
\end{figure}

To mitigate these inefficiencies, we design a merge-split-merge (MSM) dataflow that interleaves serial refinement with parallel generation, as illustrated in Fig.~\ref{fig:merge}. MSM execution begins after the FLDD-selected anchor is committed using the preceding forward pass's prediction, so that local refinement is conditioned on the updated anchor.

\begin{algorithm}[t]
\caption{FLDD-Guided Local Refinement with MSM}
\label{alg:fldd_msm_compact}
\begin{algorithmic}[1]
\Require Current committed state $X^{(r)}$;
\Statex\hspace{\algorithmicindent}retained logits $Z^{\mathrm{prev}}$; history $\Sigma^{\mathrm{prev}}$
\Ensure State $X^{(r+1)}$; logits $Z^{\mathrm{new}}$
\State $Y\gets\Call{PredictTokens}{Z^{\mathrm{prev}}}$
\State $(i,j)\gets\Call{FLDD}{Z^{\mathrm{prev}},\Sigma^{\mathrm{prev}},X^{(r)}}$
\State $\widetilde X\gets X^{(r)}$; $S\gets\varnothing$
\State $H_B\gets\varnothing$
\If{$(i,j)\ne\varnothing$}
    \State $\widetilde X_i\gets Y_i$; $S\gets\{i,j\}$
    \State $H_B\gets\Call{Embed}{\widetilde X_S}$
\EndIf
\State $H_A\gets\Call{Embed}{X^{(r)}}$
\For{each Transformer layer $\ell$}
    \State $(Q,K,V)\gets\Call{MergedQKV}_{\ell}(H_A,H_B)$
    \State $(U_A,U_B)\gets\Call{SplitAttention}_{\ell}(Q,K,V,S)$
    \State $(H_A,H_B)\gets\Call{MergedFFN}_{\ell}(U_A,U_B)$
\EndFor
\State $Z^{\mathrm{new}}\gets\operatorname{LMHead}(H_A)$
\If{$S\ne\varnothing$}
    \State $Z^{\mathrm{new}}_j\gets\operatorname{LMHead}(H_{B,j})$
\EndIf
\State $Y\gets\Call{PredictTokens}{Z^{\mathrm{new}}}$
\State $X^{(r+1)}\gets\Call{CommitUpdate}{\widetilde X,Y}$ \Comment{Keep anchor}
\end{algorithmic}
\end{algorithm}

\textit{Merge Stage}: The $N$ parallel tokens are merged with the updated anchor and its neighbor-refinement token into an $(N+2)$-token batch for QKV projection.

\textit{Split Stage}: Attention A processes the original parallel batch using the pre-anchor context. Attention B processes the updated anchor and its neighbor, reusing the parallel path's KV context with the anchor entries replaced by the updated KV produced by the current layer's merged QKV projection. The updated anchor is propagated through successive layers to provide layer-wise conditioning for neighbor refinement. In addition, GEMV operations in Attention B are executed on the reconfigurable PE array, which dynamically adapts its computation dimensions to maximize the utilization of the GEMV workload.

\textit{Merge Stage}: The outputs of Attention A and Attention B are merged into an $(N+2)$-token batch for FFN computation.

Algorithm~\ref{alg:fldd_msm_compact} summarizes FLDD-guided anchor selection and local refinement using the MSM dataflow. In Algorithm~\ref{alg:fldd_msm_compact}, $H_B$ contains the updated anchor and its neighbor-refinement token. Only the neighbor's output replaces the corresponding parallel prediction. By merging serial refinement into the parallel generation pipeline, the MSM architecture mitigates pipeline stalls while enabling efficient execution of GEMV operations through dimension-adaptive PE reconfiguration.

\section{Evaluation}
\label{sec:evaluation}

\subsection{Evaluation Methodology}
\label{sec:eval_methodology}

\subsubsection{Hardware Baselines}

We compare DynaTE against three classes of hardware platforms. First, NVIDIA AGX Orin serves as a commercial GPU baseline for low-power edge inference. Second, we construct Fixed-SA, a conventional dLLM accelerator
built around a fixed-shape systolic array~\cite{genc2021gemmini,yi2024opengemm}, with the same technology,
frequency, compute capacity, on-chip storage, and off-chip bandwidth
as DynaTE. Finally, we select dLLM-OPU and DART as representative dLLM accelerators and normalize their matrix-compute resources to the same equivalent MAC count as DynaTE for fair comparison.

\subsubsection{DynaTE Implementation}

DynaTE employs a dimension-reconfigurable PE array composed of
eight $64\times4$ subarrays, providing a total of 2,048 splittable
MAC units. Implemented in a TSMC 28-nm technology at 1\,GHz,
the array supports $64\times32$, $128\times16$, $256\times8$,
and $512\times4$ configurations. The on-chip memory hierarchy
includes double-buffered weight and QKV tile buffers, a ping-pong
activation buffer, and an attention buffer for MSM execution.
The streaming vocabulary-processing engine contains a 32-bank,
four-row elastic buffer, an eight-entry survivor queue per token
context, and two five-stage tournament cores shared by 32 token
contexts. DynaTE is connected to an HBM2 memory system with
512\,GB/s peak bandwidth.

Table~\ref{tab:hardware_config} summarizes the complete hardware
configuration. The datapath and control logic are synthesized using
Synopsys Design Compiler. On-chip SRAM area and energy are modeled
using CACTI~7.0~\cite{balasubramonian2017cacti}. For HBM modeling, we use Ramulator with HBM2 settings~\cite{kim2015ramulator}. A cycle-accurate simulator models PE execution,
on-chip and off-chip memory accesses, MSM refinement, and streaming
vocabulary processing.

\begin{table}[t]
\centering
\caption{DynaTE hardware configuration.}
\label{tab:hardware_config}
\small
\setlength{\tabcolsep}{5pt}
\renewcommand{\arraystretch}{1.08}
\begin{tabular}{l|l}
\hline
\textbf{Parameter} & \textbf{Configuration} \\
\hline
Technology / Frequency
    & TSMC 28\,nm / 1\,GHz \\

PE Total MACs
    & 2,048 \\

Array configurations
    & $64\times32$--$512\times4$ \\

Weight--activation precision
    & W8A8 or W8A4 \\

Activation buffer
    & $2\times128$\,KB \\

Weight buffer
    & $2\times64$\,KB, double-buffered \\

QKV tile buffer
    & $2\times32$\,KB, double-buffered \\

Attention buffer
    & 128\,KB \\

Elastic buffer
    & 32 banks $\times$ 4 rows, ping-pong \\

Survivor buffer
    & 8 entries/context \\

Tournament core
    & 2, five-stage pipeline \\

Top-32 tables
    & 32 entries $\times$ 32 contexts \\

Off-chip memory
    & HBM2, 512\,GB/s \\
\hline
\end{tabular}
\end{table}

\subsubsection{Models and Benchmarks}

We evaluate DynaTE on LLaDA-8B~\cite{nie2025large} and Dream-7B~\cite{ye2025dream}, two
representative diffusion language models. Both models use part-wise
decoding with confidence-based early unmasking
and approximate KV-cache reuse. The part size is set to 32, and
each part performs at most 32 denoising iterations, terminating
early once all tokens in the part have been decoded.
For masked-token pruning candidate selection, the confidence and
distance thresholds are set to $\tau_C=0.75$ and $\tau_D=3$
for both models, while the entropy threshold $\tau_H$ is set to
0.18 for LLaDA-8B and 0.21 for Dream-7B.
The FLDD weights are $(\alpha_1,\alpha_2,\alpha_3)=(0.25,0.45,0.30)$,
and the FLDD score threshold $\tau_{\mathrm{FLDD}}$ is set to 0.40
for LLaDA-8B and 0.35 for Dream-7B.
For both models, Attn-O and MLP-Down use W8A4 throughout decoding. Attn-QKV uses W8A4 in Layers 0-4 and 6, and switches from W8A8 to W8A4 in Layers 5 and 7 once the unmasked-token ratio reaches 50\%. All other activations remain W8A8.

Our benchmark suite includes GSM8K~\cite{cobbe2021training} and
MATH~\cite{hendrycksmath2021} for mathematical reasoning,
HumanEval~\cite{chen2021evaluating} and MBPP~\cite{austin2021program}
for code generation, and MMLU~\cite{hendryckstest2021}
for general knowledge. Algorithmic effectiveness and generation
quality are evaluated on these five benchmarks.

\subsection{Design Space Exploration}
\label{sec:dse}

\subsubsection{Survivor Buffer Depth}

\begin{figure}[!t]
  \centering
  \includegraphics[width=\linewidth]
    {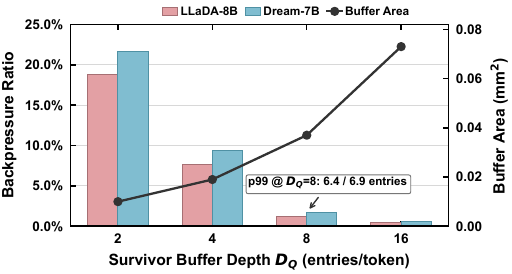}
  \caption{Survivor-buffer depth exploration.}
  \label{fig:buffer_dse}
\end{figure}

The survivor queues absorb the bursty candidates admitted by
threshold filtering. Fig.~\ref{fig:buffer_dse} reports both
buffer-full backpressure and buffer area as the per-token queue depth
increases. The backpressure ratio drops rapidly from two to eight
entries per token for both models, whereas increasing the depth to
16 yields little additional benefit while nearly doubling the buffer
area.

\subsubsection{Tournament Core Count}

\begin{figure}[!t]
  \centering
  \includegraphics[width=\linewidth]
    {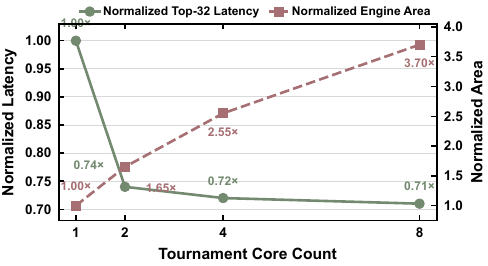}
  \caption{Tournament-core count exploration.}
  \label{fig:core_dse}
\end{figure}

Fig.~\ref{fig:core_dse} explores the number of shared five-stage
tournament cores. Increasing the core count from one to two sharply
reduces Top-32 latency, whereas moving from two to four and eight
cores provides little additional benefit while the engine area
continues to increase. We therefore share two tournament cores across
the active token contexts.

\subsection{Algorithmic Effectiveness}
\label{sec:algorithm_eval}

\begin{figure}[!t]
  \centering
  \includegraphics[width=\linewidth]
    {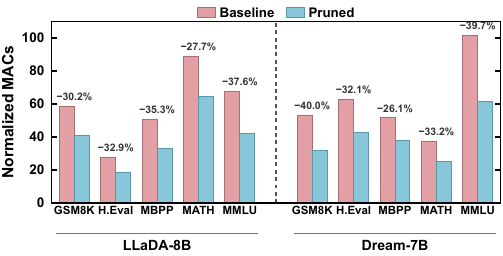}
  \caption{Token-pruning effectiveness across LLaDA-8B and Dream-7B:
  relative total Transformer MAC count before and after pruning.}
  \label{fig:prun_eval}
\end{figure}

\subsubsection{Token-Pruning Effectiveness}

We compare DynaTE with a baseline using the same part-wise decoding
configuration without token pruning, which executes all active tokens
throughout each Transformer layer.
As shown in Fig.~\ref{fig:prun_eval}, token-wise
pruning reduces the total executed Transformer MACs by
26.1--40.0\% across the evaluated workloads, with average
reductions of 32.7\% for LLaDA-8B and 34.2\% for Dream-7B,
respectively.
The reduction is larger on workloads with more low-utility
token updates, while attention-mass verification prevents
influential tokens from being pruned.

\subsubsection{FLDD Effectiveness}

\begin{figure}[!t]
  \centering
  \includegraphics[width=\linewidth]
    {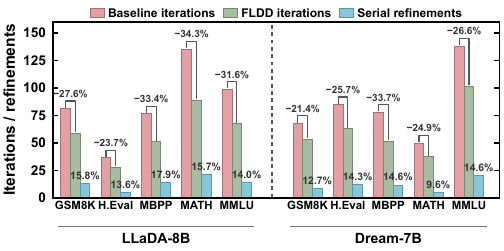}
  \caption{FLDD effectiveness across LLaDA-8B and Dream-7B.}
  \label{fig:fldd_eval}
\end{figure}

Fig.~\ref{fig:fldd_eval} compares the total number
of denoising iterations with and without FLDD-guided local
refinement.
FLDD reduces the iteration count by 21.4--34.3\% across the five
benchmarks, corresponding to average reductions of 30.1\% for
LLaDA-8B and 26.5\% for Dream-7B.
Fig.~\ref{fig:fldd_eval} also reports the local serial refinements
introduced by FLDD, averaging only 14.3\% of the baseline denoising
iterations. FLDD thus uses a small number of token-level serial
refinements to avoid a larger number of block-wide iterations.
With support from the MSM dataflow, these refinements incur only
an 8.54\% per-iteration latency overhead, as evaluated in
Section~\ref{sec:hardware_eval}.

\begin{figure}[!t]
  \centering
  \includegraphics[width=\linewidth]{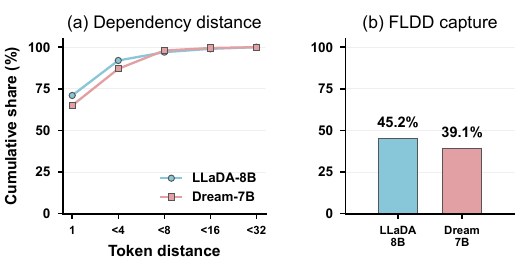}
  \caption{Dependency locality and FLDD capture: (a) cumulative
  distribution of absolute token distances among oracle dependency
  pairs; (b) fraction of the same oracle pairs captured by
  FLDD.}
  \label{fig:fldd_dependency_analysis}
\end{figure}

As shown in Fig.~\ref{fig:fldd_dependency_analysis}(a), adjacent
tokens account for 71\% and 65\% of oracle dependency pairs on
LLaDA-8B and Dream-7B, respectively, motivating refinement within
the immediate neighborhood. Fig.~\ref{fig:fldd_dependency_analysis}(b)
shows that FLDD captures 45.2\% and 39.1\% of these oracle pairs,
respectively. FLDD is therefore intended as a low-overhead heuristic
that captures useful local refinement opportunities without requiring
exhaustive coverage of token dependencies.

\subsubsection{Generation Quality}

Table~\ref{tab:ablation_accuracy} reports the generation quality
after applying each optimization. Across both models and all
benchmarks, token pruning causes a maximum score reduction of
0.6 points, while FLDD changes the scores by at most 0.5 points
and improves several workloads by up to 0.5 points. Adaptive
quantization introduces a maximum reduction of 0.3 points.
With all optimizations enabled, the average and maximum score
reductions are only 0.41 and 0.7 points, respectively, demonstrating
that DynaTE preserves generation quality.

\begin{table}[!t]
  \caption{Ablation study on generation quality. Each entry reports
  LLaDA-8B / Dream-7B.}
  \label{tab:ablation_accuracy}
  \centering
  \setlength{\tabcolsep}{2.7pt}
  \renewcommand{\arraystretch}{1.08}
  \resizebox{\columnwidth}{!}{%
  \begin{tabular}{lccccc}
    \toprule
    \textbf{Method} & \textbf{GSM8K} & \textbf{H.Eval} &
    \textbf{MBPP} & \textbf{MATH} & \textbf{MMLU} \\
    \midrule
    Baseline
      & 70.3 / 74.2
      & 35.4 / 39.5
      & 40.0 / 44.3
      & 31.4 / 29.1
      & 43.9 / 42.2 \\

    + Pruning
      & 69.9 / 74.1
      & 35.2 / 39.3
      & 39.7 / 44.5
      & 30.8 / 28.8
      & 43.6 / 41.6 \\

    + FLDD
      & 70.4 / 74.2
      & 35.3 / 39.4
      & 40.0 / 44.1
      & 31.9 / 29.2
      & 43.8 / 42.0 \\

    + Quant.
      & 70.5 / 73.9
      & 35.2 / 39.2
      & 39.7 / 44.1
      & 31.3 / 28.9
      & 43.8 / 42.3 \\

    \textbf{+ All}
      & \textbf{69.8 / 73.9}
      & \textbf{34.9 / 39.4}
      & \textbf{39.3 / 43.7}
      & \textbf{30.8 / 28.9}
      & \textbf{43.7 / 41.8} \\
    \bottomrule
  \end{tabular}}
\end{table}

\subsection{Hardware Evaluation}
\label{sec:hardware_eval}

\subsubsection{End-to-End Performance and Energy}

\begin{figure*}[!t]
  \centering
  \includegraphics[width=\textwidth]
  {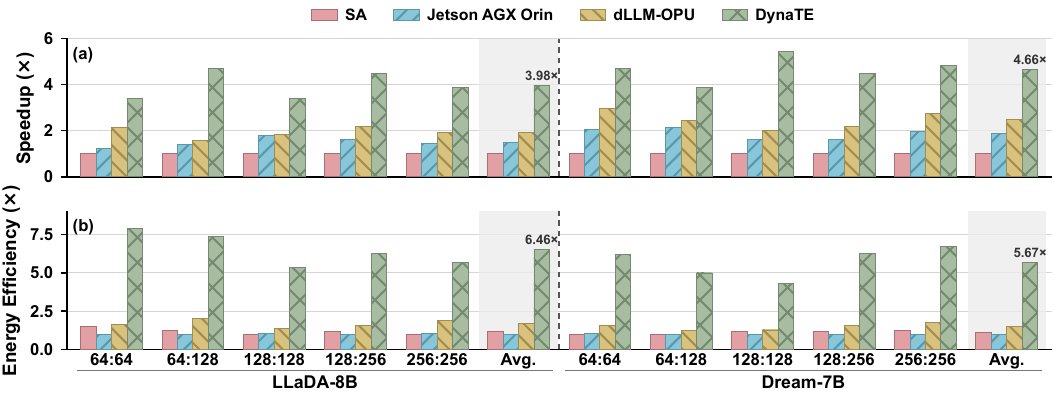}
  \caption{End-to-end comparison across LLaDA-8B and Dream-7B:
  (a) speedup and (b) energy-efficiency improvement.}
  \label{fig:end_to_end}
\end{figure*}

Fig.~\ref{fig:end_to_end} compares SA, Jetson AGX Orin, dLLM-OPU,
and DynaTE on LLaDA-8B and Dream-7B under different
input/output sequence lengths. All four configurations enable adaptive
decoding with early stopping, allowing decoding to terminate before
the maximum iteration budget rather than enforcing a fixed number of
iterations. Jetson AGX Orin and SA execute the same baseline decoding algorithm without additional algorithmic optimizations, while DynaTE and dLLM-OPU apply their respective workload-specific optimizations. DynaTE achieves average speedups of 2.55$\times$,
4.32$\times$, and 1.96$\times$ over Jetson AGX Orin, SA, and
dLLM-OPU, respectively, together with energy-efficiency
improvements of 6.07$\times$, 5.26$\times$, and 3.82$\times$.

\subsubsection{Reconfigurable-Array Effectiveness}

\begin{figure}[!t]
  \centering
  \includegraphics[width=\linewidth]
    {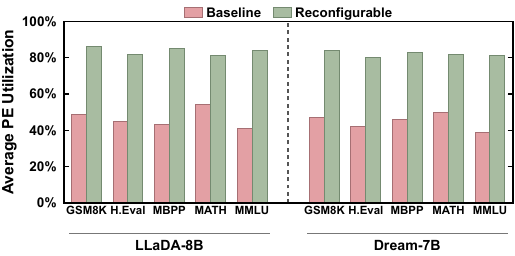}
  \caption{Average PE utilization of the fixed and reconfigurable
  arrays across LLaDA-8B and Dream-7B workloads.}
  \label{fig:pe_utilization}
\end{figure}

The irregular operator shapes produced by token pruning can
underutilize a fixed array. Fig.~\ref{fig:pe_utilization} reports
the resulting hardware utilization. Across both models and all five
workloads, DynaTE's reconfigurable array maintains over 80\% average
PE utilization, substantially higher than the matched fixed-array
baseline.

\begin{figure}[!t]
  \centering
  \includegraphics[width=\linewidth]
    {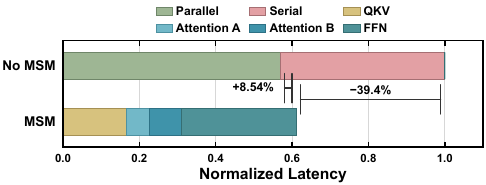}
  \caption{Normalized per-iteration latency breakdown with and
  without the MSM dataflow.}
  \label{fig:msm_eval}
\end{figure}

\subsubsection{MSM Dataflow Effectiveness}

Fig.~\ref{fig:msm_eval} shows how MSM supports the serial refinement
introduced by FLDD. Independent refinement turns QKV and FFN into
low-utilization GEMV execution. MSM instead merges the serial token
with the parallel batch for QKV and FFN, hiding much of the additional
serial computation and reducing per-iteration latency by 39.4\%
compared with independently executed serial refinement without MSM.
With MSM, local refinement requires only an 8.54\% increase in
per-iteration latency relative to parallel-only execution, enabling
a reduction in the total number of denoising iterations at a modest
additional cost per iteration.

\begin{table}[!t]
  \caption{Area and power breakdown of DynaTE.}
  \label{tab:area_power}
  \centering
  \setlength{\tabcolsep}{4pt}
  \renewcommand{\arraystretch}{1.08}
  \begin{tabular}{lcccc}
    \toprule
    \textbf{Metric} &
    \textbf{Logic} &
    \textbf{SRAM} &
    \textbf{HBM} &
    \textbf{Overall} \\
    \midrule
    Area (mm$^2$)
      & 2.12 & 1.79 & -- & 3.91 \\
    Power (W)
      & 1.01 & 1.18 & 5.17 & 7.36 \\
    \bottomrule
  \end{tabular}
  \par\medskip
  \includegraphics[width=\linewidth]{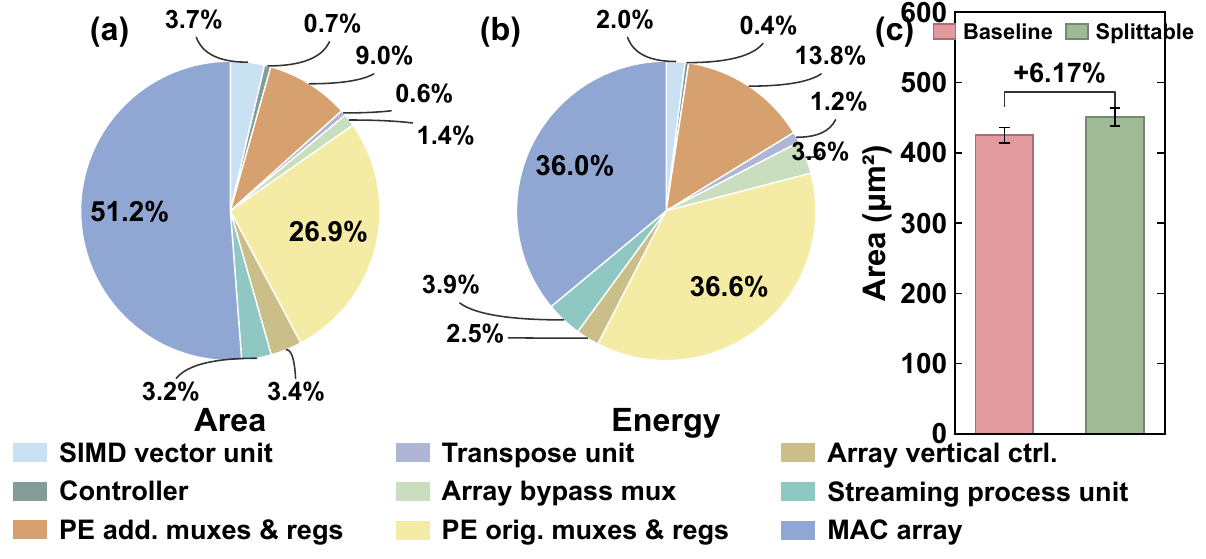}
  \captionof{figure}{Hardware cost: (a)(b) logic-area and logic energy
  breakdown, and (c) area comparison between the baseline and
  splittable MAC units.}
  \label{fig:hardware}
\end{table}

\subsubsection{Area and Energy Breakdown}

Table~\ref{tab:area_power} and Fig.~\ref{fig:hardware} report the area and energy
breakdowns of DynaTE. The MAC array and the original PE datapath
remain the dominant contributors. The additional routing and control
logic required for reconfiguration forms a modest fraction of the
overall design. The streaming process unit contributes only 3.2\%
of the logic area and 3.9\% of the logic energy. Fig.~\ref{fig:hardware}(c) compares the proposed
splittable MAC with the baseline MAC. Supporting multiple precision
modes increases the MAC area by only 6.17\%, while iteration-adaptive
precision reduces the PE-array energy by 32.6\% compared with
uniform W8A8 execution.

\subsubsection{Streaming Vocabulary Processing}

\begin{figure}[!t]
  \centering
  \includegraphics[width=\linewidth]
    {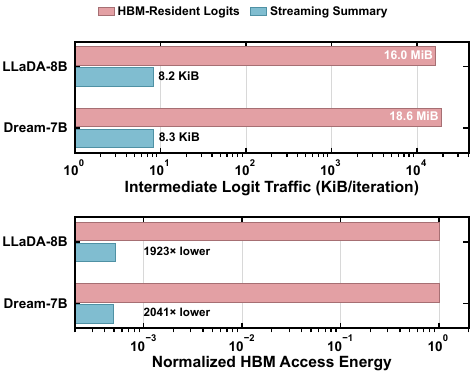}
  \caption{Intermediate-logit traffic and normalized HBM access
  energy for HBM-resident logits and streaming summaries.}
  \label{fig:streaming_gain}
  \vspace{-2mm}
\end{figure}

Fig.~\ref{fig:streaming_gain} compares the conventional
HBM-resident logit pipeline with DynaTE's streaming engine.
For LLaDA-8B, DynaTE reduces intermediate-logit traffic from
16.0\,MiB to 8.2\,KiB per iteration and lowers the corresponding
HBM access energy by 1,923$\times$. For Dream-7B, the traffic
is reduced from 18.6\,MiB to 8.3\,KiB, yielding a
2,041$\times$ reduction in HBM access energy.

\subsubsection{Multi-token Scheduler Effectiveness}

\begin{figure}[!t]
  \centering
  \includegraphics[width=\linewidth]
    {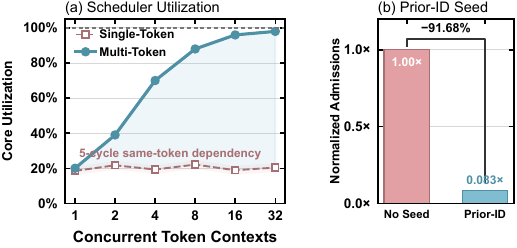}
  \caption{Top-32 scheduling and filtering effectiveness:
  (a) tournament-core utilization under single-token and multi-token
  scheduling; (b) normalized logits admitted to the tournament tree
  without and with Prior-ID seeding. }
  \label{fig:top32_utilization}
  \vspace{-2mm}
\end{figure}

A single token can issue only one tournament update every five cycles
because of the pipeline dependency. Fig.~\ref{fig:top32_utilization}
shows that single-token scheduling therefore fluctuates around 20\%
core utilization. Interleaving independent token contexts fills the
pipeline bubbles and raises utilization toward 97.6\% as the number of
concurrent contexts increases. Prior-ID seeding further reduces the
number of logits admitted to the tournament tree by 91.68\%.

\subsubsection{Overall Ablation}

\begin{figure}[!t]
  \centering
  \includegraphics[width=\linewidth]
    {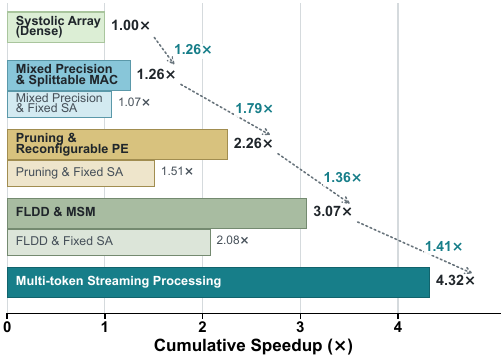}
  \caption{Incremental ablation of DynaTE.}
  \label{fig:overall_ablation}
  \vspace{-2mm}
\end{figure}

Fig.~\ref{fig:overall_ablation} shows the cumulative speedup
from DynaTE's main optimizations. Mixed precision improves the
baseline to 1.26$\times$, followed by 2.26$\times$ with token
pruning, 3.07$\times$ with FLDD, and 4.32$\times$ with streaming
vocabulary processing. We also evaluate the corresponding algorithmic
optimizations on a fixed systolic array: the mixed-precision,
token-pruning, and FLDD configurations achieve cumulative speedups
of 1.07$\times$, 1.51$\times$, and 2.08$\times$, respectively,
over the same dense baseline. The lower gains in these configurations
highlight the importance of coupling the algorithmic optimizations
with the splittable MAC datapath, reconfigurable array, and MSM dataflow
to translate execution changes into higher end-to-end performance.

\subsubsection{Comparison with Prior dLLM Accelerators}

Table~\ref{tab:accelerator_comparison} reports the average
results across LLaDA-8B and Dream-7B on the five evaluated
benchmarks, with the generation length fixed at 256 tokens.
DART, SA, and DynaTE use the same HBM configuration,
while dLLM-OPU retains its native FPGA memory system and
decoding pipeline. Compared to three accelerators, DynaTE improves speed and energy efficiency by 2.05--4.32$\times$ and 2.99--5.28$\times$,
respectively.

\begin{table}[!t]
  \caption{Comparison with prior dLLM accelerators.}
  \label{tab:accelerator_comparison}
  \centering
  \scriptsize
  \setlength{\tabcolsep}{2.2pt}
  \renewcommand{\arraystretch}{1.08}
  \resizebox{\columnwidth}{!}{%
  \begin{tabular}{lcccc}
    \toprule
    \textbf{Metric}
      & \textbf{DART}
      & \textbf{dLLM-OPU}
      & \textbf{SA}
      & \textbf{DynaTE} \\
    \midrule

    Platform
      & ASIC
      & U200 FPGA
      & ASIC
      & ASIC \\

    Technology
      & 7\,nm
      & 16\,nm
      & 28\,nm
      & 28\,nm \\

    Frequency
      & 1\,GHz
      & 300\,MHz
      & 1\,GHz
      & 1\,GHz \\

    Area (mm$^2$)
      & 3.20\textsuperscript{a}
      & --
      & 3.34
      & 3.91 \\

    \midrule

    Throughput (tokens/s)
      & 8.97
      & 12.12
      & 5.76
      & 24.89 \\

    Area eff. (tokens/s/mm$^2$)
      & 2.81
      & --
      & 1.72
      & 6.36 \\

    Energy eff. (tokens/J)
      & 1.13
      & 0.86
      & 0.64
      & 3.38 \\

    \bottomrule
  \end{tabular}}

  \parbox{\columnwidth}{\scriptsize
  \textsuperscript{a}Area normalized to 28\,nm using the scaling model in~\cite{stillmaker2017scaling}.}
\end{table}

\section{Related Work}

\textbf{dLLM Accelerators.}
Only a few recent accelerators have begun to address the distinct
sampling and computation-reuse patterns of dLLMs. DART~\cite{lou2026npudesigndiffusionlanguage} accelerates diffusion sampling with dedicated hardware for vocabulary-wide confidence computation and token selection. However, the logits produced by the LM head are first written to HBM and then read back in chunks for sampling. DynaTE instead processes logits directly as they emerge from the LM head, avoiding intermediate full-logit HBM transfers, and interleaves multiple token streams to accommodate irregular output patterns.
dLLM-OPU~\cite{wei2026dllm} reduces redundant Transformer computation
through cross-iteration reuse of Attention and FFN activations,
exploiting temporal similarity across adjacent denoising steps.
Such cross-iteration reuse requires retaining intermediate activations
from previous denoising steps, introducing additional storage requirements.
DynaTE focuses on refinement utility and local dependencies to adapt
token participation and selectively perform local refinement within
an iteration, with corresponding hardware support for the resulting
variable workloads.

\textbf{dLLM Algorithmic Acceleration.}
SureLock~\cite{oba2026surelock} reduces MAC operations by locking
unmasked tokens with low KL divergence between full-vocabulary
posterior distributions across consecutive iterations and reusing
their cached KV states thereafter. DynaTE instead targets masked
tokens with low expected refinement utility and makes layer-wise
skipping decisions based on received attention mass. Attention mass is
accumulated directly from the attention stream, while candidate
selection reuses existing distribution statistics.
DynaTE further employs a dimension-reconfigurable PE array to
accommodate varying active-token counts and maintain utilization
after pruning.

\textbf{DNN Hardware Optimizations.}
For array reconfiguration, Planaria~\cite{ghodrati2020planaria} dynamically partitions accelerator
resources to support spatial multi-tenant DNN execution, whereas
DynaTE focuses on dimension reconfiguration driven by token pruning
and local serial refinement in dLLM inference.
ReDas~\cite{han2024redas} supports fine-grained reshaping through
roundabout datapaths and four surrounding multi-mode buffers.
DynaTE uses one-sided SRAM bypass paths and inter-subarray vertical
connections to implement simpler data routing for token-dimension
reconfiguration.
For precision reconfiguration, BitFusion~\cite{sharma2018bitfusion}
uses bit-level fusion to support variable-precision multiplication
while retaining a fixed 32-bit partial-sum datapath.
DynaTE reconfigures shared adder segments into one 24-bit accumulation
path for W8A8 or two independent 20-bit paths for dual-W8A4,
utilizing all constituent multipliers and adder segments in both modes.
For Top-K selection, SpAtten~\cite{wang2021spatten} uses pivot-based
partitioning and parallel zero elimination to accelerate selection
within a single input vector. DynaTE instead interleaves vocabulary
candidates from multiple token streams through shared tournament
pipelines to accommodate uneven selection workloads across tokens.

\section{Conclusion}

We presented DynaTE, an iteration-aware accelerator that restructures
dLLM inference around dynamic token execution. Token-wise pruning
and FLDD-guided refinement reduce redundant computation and
denoising iterations, while a reconfigurable PE array, splittable
MAC units, and MSM dataflow efficiently execute the resulting
dynamic workloads. A streaming multi-token vocabulary engine further
eliminates full-logit HBM transfers and reuses vocabulary statistics
for pruning, dependency detection, and sampling. Evaluations on two representative dLLMs and five benchmarks show
that DynaTE achieves speedups of 2.05--2.78$\times$ and
energy-efficiency improvements of 2.99--3.93$\times$ over
state-of-the-art dLLM accelerators, while delivering a
2.55$\times$ speedup and 6.07$\times$ higher energy efficiency
than Jetson AGX Orin.

\section*{Acknowledgment}
This work was supported by the National Natural Science Foundation of China (Grant No. 62274142) and the Zhejiang Province's Leading Talent Project in Science and Technology Innovation (2023R5204). AI tools (ChatGPT and Claude) were used only to polish the language and improve the readability of this paper.

\bibliographystyle{IEEEtran}
\bibliography{ref}

\end{document}